\documentclass[prl,twocolumn,nofootinbib]{revtex4}
\usepackage{graphicx}
\usepackage{latexsym}
\def\be{\begin{equation}}
\def\ee{\end{equation}}
\def\bea{\begin{eqnarray}}
\def\eea{\end{eqnarray}}
\def\btheta{\tilde{\theta}}

\begin{document}
\title{An inflation model with large variations in spectral index}
\author{Bo Feng${}^a$}
\author{Mingzhe Li${}^a$}
\author{Ren-Jie Zhang${}^{b,c}$}
\author{Xinmin Zhang${}^a$}
\affiliation{${}^a$Institute of High Energy Physics, Chinese
Academy of Science, P.O. Box 918-4, Beijing 100039, P. R. China}
\affiliation{${}^b$School of Natural Sciences, Institute for
Advanced Study, Princeton, NJ 08540, USA}
\affiliation{${}^c$Institute of Theoretical Physics, Chinese
Academy of Sciences, Beijing 100080, P. R. China.}

\begin{abstract}
Recent fits of cosmological parameters by the Wilkinson Microwave
Anisotropy Probe (WMAP) measurement favor a primordial scalar
spectrum with varying index. This result, if stands, could
severely constrain inflation model buildings. Most extant
slow-roll inflation models allow for only a tiny amount of scale
variations in the spectrum. We propose in this paper an
extra-dimensional inflation model which is natural theoretically
and can generate the required variations of the
spectral index as implied by the WMAP for suitable choices of parameters.\\
PACS number(s): 98.80.Cq, 11.10.Kk
\end{abstract}

\maketitle

The recently released Wilkinson Microwave Anisotropy Probe (WMAP)
data \cite{WMAP} have been used to fit the cosmological parameters
and confront the predictions of inflationary scenarios
respectively in Refs.~\onlinecite{WMAP1} and \onlinecite{WMAP2}.
It is found that the data (with no other significant priors) can
be best fitted by a standard $\Lambda$CDM model seeded by an
almost scale-invariant, adiabatic, Gaussian primordial
fluctuation; predictions inferred from the model also agree with
other cosmological measurements with high
accuracies\cite{melchiorri}. It is noted that there might be
possible discrepancies between predictions and observations on the
largest and smallest scales. The problem at the smallest scales is
improved when combined with data from finer scale CMB experiments
(ACBAR and CBI) and structure formation measurements (2dFGRS and
Lyman $\alpha$ forest), for which a varying scalar primordial
spectrum is favored by the best fit. At the pivot scale
$k_0=0.05~{\rm Mpc}^{-1}$, the best-fit values for the scalar
power spectrum are $n_s=0.93\pm 0.03$ and $dn_s/d\ln
k=-0.031^{+0.016}_{-0.018}$ \cite{WMAP1}; this means at a
$2\sigma$ level the spectrum runs from blue to red as the
co-moving wave number $k$ increases. As noted, the suppression of
the power at the smallest scale might offer an interesting
solution to the problem of the standard $\Lambda$CDM model at
small scales \cite{SS}.

The intriguing result of a varying spectral index needs to be
closely evaluated with improved statistics and with future data.
If stands, it could severely constrain inflation model buildings.
Most extant inflationary models allow for only a tiny amount of
scale variations.  In this paper we propose a single-field
inflation model which is natural theoretically and can generate
the required variations of the spectral index as implied by the
WMAP.

For the single-field slow-roll inflationary models, one usually defines the
following slow-roll parameters \cite{Liddle2},
\be\label{slowroll}
\epsilon \equiv\frac{M_P^2}{2}\left(\frac{V'}{V}\right)^2~,~~
\eta \equiv M_P^2\frac{V''}{V}~,~~
\xi \equiv M_P^4\frac{V'V'''}{V^2}~,
\ee
where primes represent derivatives with respect to the inflaton field, $\phi$,
and $M_P=2.4\times 10^{18}$ GeV is the reduced Planck
mass. Slow-roll approximation requires $\epsilon,~|\eta|,~|\xi|\ll 1$
in the inflationary epoch.

The primordial curvature (scalar) and tensor perturbation
power spectra are given by \cite{Liddle2}
\be
{\cal P}_{\cal R}\approx \left.\frac{V}{24\pi^2M_P^4\epsilon}\right|_{k=aH}~,\quad
{\cal P}_h \approx \left.\frac{2V}{3\pi^2 M_P^4}\right|_{k=aH}~,
\ee
evaluated when a particular mode crosses out the horizon.
The tensor to scalar ratio $r\equiv{\cal P}_h/{\cal P}_{\cal R}\approx 16\epsilon$ is generally small in the slow-roll inflation.
The spectral indices and their running are the slopes and curvatures of the power spectra.
In terms of the slow-roll parameters, for scalar perturbations, they are
\be
n_s-1 \approx -6\epsilon+2\eta,~~\quad
\frac{dn_s}{d\ln k}\approx 16\epsilon\eta-24\epsilon^2-2\xi.~~
\ee
In most inflationary models, $\epsilon, |\eta|\sim
M_P^2/(\Delta\phi)^2$, $|\xi|\sim M_P^4/(\Delta\phi)^4$, and
the number of $e$-folding ${\cal N}\sim (\Delta\phi)^2/M_P^2$, where $\Delta\phi$ is the displacement of the
homogeneous field $\phi$, so there is a hierarchy in the slow-roll parameters,
$\epsilon,~|\eta|\sim {\cal N}^{-1}$, $|\xi|\sim {\cal N}^{-2}$, and
the variations of the spectral indices, $dn_s/d\ln k$, are negligible.

To achieve a running spectrum in the order favored by the WMAP, one
has to consider inflation models with more exotic potentials \cite{dodelson}, such as the running mass
model \cite{stewart} and models with oscillating primordial spectrum \cite{wang}.
These models can generate significant spectral runnings, which could
be as large as $n_s-1$.

The Achilles heel of these inflationary models (and of the
inflation paradigm in general) lies, arguably, in the difficulty
of obtaining a sufficiently flat and stable (against radiative
corrections) inflaton potential from the perspectives of particle
physics. Symmetry principles must be invoked. There are only two
known symmetries which can protect the flatness of a scalar
potential: supersymmetry and the shift symmetry for a Pseudo
Nambu-Goldstone Boson (PNGB). However, as shown in
Ref.~\onlinecite{copeland} and recently re-emphasized in
Refs.~\onlinecite{ACCR} and \onlinecite{KW}, supersymmetry alone
cannot protect the flatness of the inflaton potential, since it is
explicitly broken during inflation, and the gravitational effects
generically give a Hubble-scale mass correction to the inflaton.

The shift symmetry was first realized in the natural
inflation model \cite{Freese}. Still there are some difficulties in this
model. The flatness condition, in the simplest scenario with a single
PNGB, requires the scale of spontaneous symmetry breaking
and the values of the inflaton during the slow roll above
the Planck scale, taking the model outside the regime of
validity of an effective field theory description. Moreover, it is
expected that the gravity-induced higher-dimensional
operators are not suppressed.

These issues have been recently re-examined in the context of
extra dimensions (called extra-natural inflation in
Ref.~\onlinecite{ACCR}). Consider a five-dimensional Abelian gauge
field model with the fifth dimension compactified on a circle of
radius $R$, we identify the inflaton field $\theta$ with the
gauge-invariant Wilson loop of the extra component $A_5$
propagating in the bulk, \be \theta=g^{}_5\oint dx^5 A_5~, \ee
where $g_5$ is the five-dimensional gauge coupling constant. At
energies below $1/R$, $\theta$ is a four-dimensional field with an
effective Lagrangian
\begin{equation}\label{Lagrangian}
{\cal L} = \frac{1}{2g_4^2 (2 \pi R)^2} (\partial \theta)^2 -
V(\theta)\;,
\end{equation}
with $g_4^2 =g_5^2/2 \pi R$ the four-dimensional effective gauge coupling
constant. The non-local potential $V(\theta)$ is generated in the presence of
particles charged under the Abelian symmetry \cite{hosotani}.

For bulk fields with bare masses $M_a$ and
charges $q_a$ the potential takes the form \cite{Delgado}
\be\label{potential}
V(\theta)=\frac{1}{128\pi^6R^4}{\rm Tr}\biggl[V(r^F_a,\theta)-V(r^B_a, \theta)\biggr]~,
\ee
where the trace is over the number
of degrees of freedom, and the superscripts $F$ and $B$ stand for
fermions and bosons respectively. Here
\bea\label{potential2}
V(r_a,\theta)&=&x_a^2\,{\rm Li}_3(r_ae^{-x_a})+3x_a\,{\rm Li}_4(r_ae^{-x_a})\nonumber\\
&+& 3\,{\rm Li}_5(r_ae^{-x_a})+h.c.~, \eea with \be
r_a=e^{iq_a\theta}\,,\quad x_a= 2\pi R M_a\,, \ee and the
poly-logarithm function ${\rm Li}_k(z)$ are \be {\rm Li}_k(z)=
\sum_{n=1}^{\infty}\frac{z^n}{n^k}~. \ee For massless particles
($x_a=0$) considered in Ref.~\onlinecite{ACCR} the potential is
\begin{equation}\label{eq:WLpot}
V(\theta) = -\frac{3}{64\pi^6R^4}
\sum_I (-)^{F_I}\sum_{n=1}^{\infty}\frac{\cos(n q \theta)}{n^5} \;,
\end{equation}
where $F_I=0$ and $1$ stand for massless bosonic and fermionic fields,
respectively.

Neglecting the higher power terms in Eq. (\ref{eq:WLpot}), one obtains
the same form of the potential as that of the natural inflation model.
The effective decay constant of the spontaneously broken Abelian symmetry is
\begin{equation} \label{eq:decay}
f_{\rm eff}  = \frac{1}{2\pi g_4 R} \;,
\end{equation}
which can be naturally greater than $M_P$ for a sufficiently small
coupling constant $g_4$ \cite{ACCR}. Moreover, due to the extra
dimension nature, gravity-induced higher-dimensional operators are
generally exponentially suppressed. This solves the forementioned
problems of the four-dimensional natural inflation
model\footnote{If natural inflation model includes a large Z in
the kinetic term\cite{largez}, redefining the field gives rise to
an effective decay constant $f_{eff}=\sqrt{Z} f$, which ( as well
as the inflaton field itself ) can also be larger than $M_p$,
however a question remained is how to get a large Z
naturally\cite{prepare}.}. The extra-natural model predicts a
red-tilted scalar spectrum with negligible spectral runnings, same
as that of the natural inflation.

The model we propose in this paper includes one massless and one
massive fields\footnote{Massive particles have also been
considered in Ref.~\onlinecite{riotto}, in the context of
extra-dimensional quintessence models.} coupled to $A_5$, {\it
i.e.}, $M_1=0,~M_2\gtrsim 1/R$, the corresponding potential for
$\theta$ is \bea
V(\theta)&=&-\frac{3}{64\pi^6R^4}\sum_{n=1}^{\infty}\frac{1}{n^3}\Biggl[(-)^{F_1}\frac{\cos
(nq_1\theta)}{n^2}\nonumber\\
&+& (-)^{F_2}
e^{-nx_2}\left(\frac{x_2^2}{3}+\frac{x_2}{n}+\frac{1}{n^2}\right)\cos
(nq_2\theta)\Biggr]\;. \eea Neglecting the higher power terms in
the sum, and defining a canonical field $\phi=f_{\rm eff}\theta$,
the effective Lagrangian of our model becomes\footnote{For a
specific presentation, we set $F_1=1$. The $F_1=0$ case is
equivalent since it only corresponds to a co-ordinate shift in the
potential.} \be\label{eqpotential} {\cal
L}=\frac{1}{2}(\partial\phi)^2-V_0\biggl[1-
\cos\left(\frac{q_1\phi}{f_{\rm eff}}\right)-\sigma \cos
\left(\frac{q_2\phi}{f_{\rm eff}}\right)\biggr]\,, \ee where \be
\sigma=(-)^{F_2+1} e^{-x_2}\left(\frac{x_2^2}{3}+x_2+1\right), ~~
V_0=\frac{3}{64\pi^6R^4}~. \ee Note that in our calculations we
have added a $\sigma$-dependent term to the potential, Eq.
(\ref{eqpotential}), to make it vanish at the minimum. For
$\sigma=0$, this potential coincides with that of the natural
inflation model.

The slow-roll parameters of Eq. (\ref{slowroll}) are \bea
\epsilon&=&\frac{\mu^2}{2}\frac{(\sin\btheta+\sigma \kappa\sin
\kappa\btheta)^2 }{[1-
\cos\btheta-\sigma\cos\kappa\btheta]^2}~,\quad\\
\eta&=&\frac{\mu^2(\cos\btheta+ \sigma \kappa^2 \cos\kappa\btheta)
}{1- \cos\btheta-\sigma \cos\kappa\btheta
}~,\label{eq:eta!}\quad\\ \xi&=&-
\frac{\mu^4(\sin\btheta+\sigma\kappa\sin\kappa\btheta)(\sin\btheta+\sigma\kappa^3
\sin\kappa\btheta)}{[1- \cos\btheta-\sigma\cos\kappa\btheta]^2}
~,\quad \label{eq:ksi!} \eea
where we have defined $\btheta\equiv q_1\theta$ and
\be
\mu\equiv q_1 M_P/f_{\rm eff}\,,\quad\kappa\equiv q_2/q_1\,.
\ee

To see analytically the effects of $\sigma$ term, we consider
the following choice of parameters: $\kappa\gg 1$, $\sigma\ll 1$,
$|\sigma| \kappa\ll 1 $ and $|\sigma|\kappa^2\sim {\cal O}(1)$.
When cosmological scales begin to cross out the
horizon in the inflationary epoch, $\btheta\sim \pi$, the
slow-roll parameters are approximately
\bea
& &\epsilon\sim \mu^2(\pi-\btheta)^2~,\\
& &\eta\sim \mu^2 (-1+\sigma\kappa^2\cos \kappa\btheta)~,\label{eta2}\\
& &\xi\sim -\mu^4\sigma\kappa^3(\pi-\btheta)\sin\kappa\btheta
~.\label{eq:ksi2}
\eea
Hence $\epsilon\ll |\eta|$.
The spectral index is determined by the $\eta$ parameter, $n_s-1\simeq 2\eta$, and the tensor fluctuation is negligible.

From the above equations, one generically would
have $|\xi|\sim \eta^2\sim 10^{-4}$, corresponding to a
negligible spectral variation.
However, the large $\sigma\kappa^3$ term  in Eq.~(\ref{eq:ksi2}) can enhance $\xi$
substantially (to the size as large as $\eta$), and lead to a
scale varying spectrum. Furthermore, for appropriate choices of $\sigma$ and $\kappa$, the
scalar power spectrum can run from blue to red as the co-moving scale $k$ increases.
Such a tilted spectrum is favored by the current data of WMAP \cite{WMAP1}.

\begin{figure}[htbp]
\includegraphics[scale=0.6]{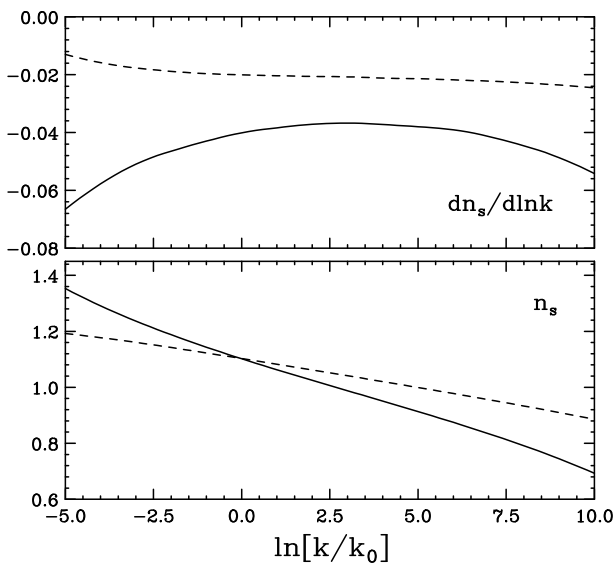}
\caption{Spectral indices $n_s$ and their runnings $dn_s/d \ln k$
for $\sigma=-3.8\times 10^{-4}$, $\mu=1/2$, $\kappa=100$ (solid
lines), and for $\sigma=0.012$, $\mu=1/3$, $\kappa=15$ (dashed
lines). $k_0=0.002~{\rm Mpc}^{-1}$ is the pivot scale of WMAP.
\label{fig:spec}}
\end{figure}

We show this behavior of power spectrum running in
Fig.~\ref{fig:spec}, for the following choices of parameters: (a)
$\sigma=-3.8\times 10^{-4}$, $\mu=1/2$ and $\kappa =100$ (solid
lines), and (b) $\sigma=0.012$, $\mu= 1/3$, $\kappa=15$ (dashed
lines). To ensure the accuracy of our results, we have retained
the higher power terms (up to $n=6$ for the massless particle) in
the numerical calculations. We have also set the number of
$e$-folding ${\cal N}(k_*)=50$ at a reference co-moving scale
$k_*$; the WMAP analyses  used pivot scales $k_0=0.002$ and $0.05$
${\rm Mpc}^{-1}$ \cite{WMAP1, WMAP2}. (This scale arbitrariness is
the inherent theoretical uncertainty in our analysis; it can be
resolved if the detail reheating history is known \cite{feng}.)
Normalizing the spectral index to the WMAP central value $n_s=1.1$
at $k_0=0.002~{\rm Mpc}^{-1}$ \cite{WMAP2}, we find $dn_s/d\ln
k\approx -0.041$ and $\approx -0.021$ for case (a) and (b)
respectively; they are in good agreements with the current WMAP
fits \cite{WMAP1, WMAP2}. Such a running feature of the spectral
index does not exist in the usual natural or extra-natural
inflation models.

\begin{figure}[htbp]
\includegraphics[scale=0.35]{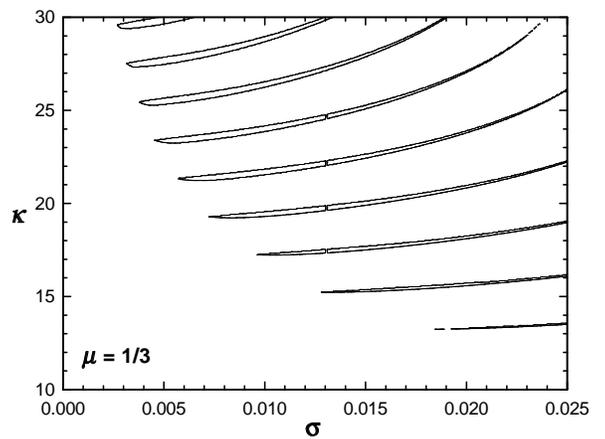}
\caption{Regions in the $\kappa$-$\sigma$ plane which give
variations of the spectral index in the range $-0.084<dn_s/d\ln
k<-0.027$, for $\mu=1/3$. The spectral indices have been
normalized to $n_s=1.13$ at the pivot scale $k_0$.
\label{fig:fig2}}
\end{figure}

In Fig.~\ref{fig:fig2} we delineate the parameter space that gives
the required amount of spectral runnings ($-0.084<dn_s/d\ln
k<-0.027$)\cite{WMAP2} in our model. We have normalized the
spectral index $n_s=1.13$ at the same pivot scale $k_0$. We show
allowed regions in the $\kappa$-$\sigma$ plane, for $\mu=1/3$. The
correlation between $\kappa$ and $\sigma$ can be understood
because one needs a certain amount of cancellations between the
two terms in Eq. (\ref{eta2}) (to achieve a spectral running from
blue to red).

In our models the size of the fifth dimension is determined to be of the order $R\sim 10/M_P$
from the COBE normalization, ${\cal P}^{1/2}_{\cal R}\sim 10^{-5}$. For
our choices of parameters, $\mu\sim 0.1$, this implies the
four- and five-dimensional gauge coupling constants $g_4\lesssim 10^{-3}/q_1$,
$g_5^2\lesssim 10^{-4}/q_1^2M_P$.

These parameter choices could be accommodated in models of
fundamental theories. For example, string theory predicts a host
of exotic particles with fractional charges; one can obtain large
charge ratios of the massive to massless particles ($\kappa\sim
10$-$100$) as required in our models. The massive particle has a
mass of the order $M_2\sim {\cal O}(1)R^{-1}$, falls in the range
of massive string states. It remains to be seen whether this kind
of model can arise naturally from specific string models.

The small gauge coupling constant ($g_4\lesssim 10^{-3}$ when
$q_1\sim {\cal O}(1)$) might seem worrisome, since string theory
compactification generally requires $g_4 M_P R\gtrsim 1$
\cite{ACCR}. This problem can be improved by introducing many
Abelian gauge fields in higher dimensions as in the assisted
inflation model \cite{liddle}. For a system with $N$ Wilson loops
$\theta_i$, the Lagrangian is given by
\begin{equation}
{\cal L} = \sum_{i=1}^{N} \frac{1}{2g_4^2 (2 \pi R)^2} (\partial
\theta_i)^2 - \sum_{i=1}^{N}  V(\theta_i) \;.
\end{equation}
 This system has a solution where all the $\theta_i$ fields are
equal, consequently it can be described by a single-field
model through the redefinitions, $\widetilde{f}_{\rm
eff}=\sqrt{N}/2\pi g_4 R$, $\widetilde{V}=NV_i$. The COBE
normalization gives $R N^{-1/4}\sim 10 M_P^{-1}$, which implies
$g_4 \lesssim 10^{-3}N^{1/4}$. The above requirement can be naturally satisfied if $N\gtrsim 100$.

\begin{figure}[htbp]
\begin{center}
\includegraphics[scale=0.35]{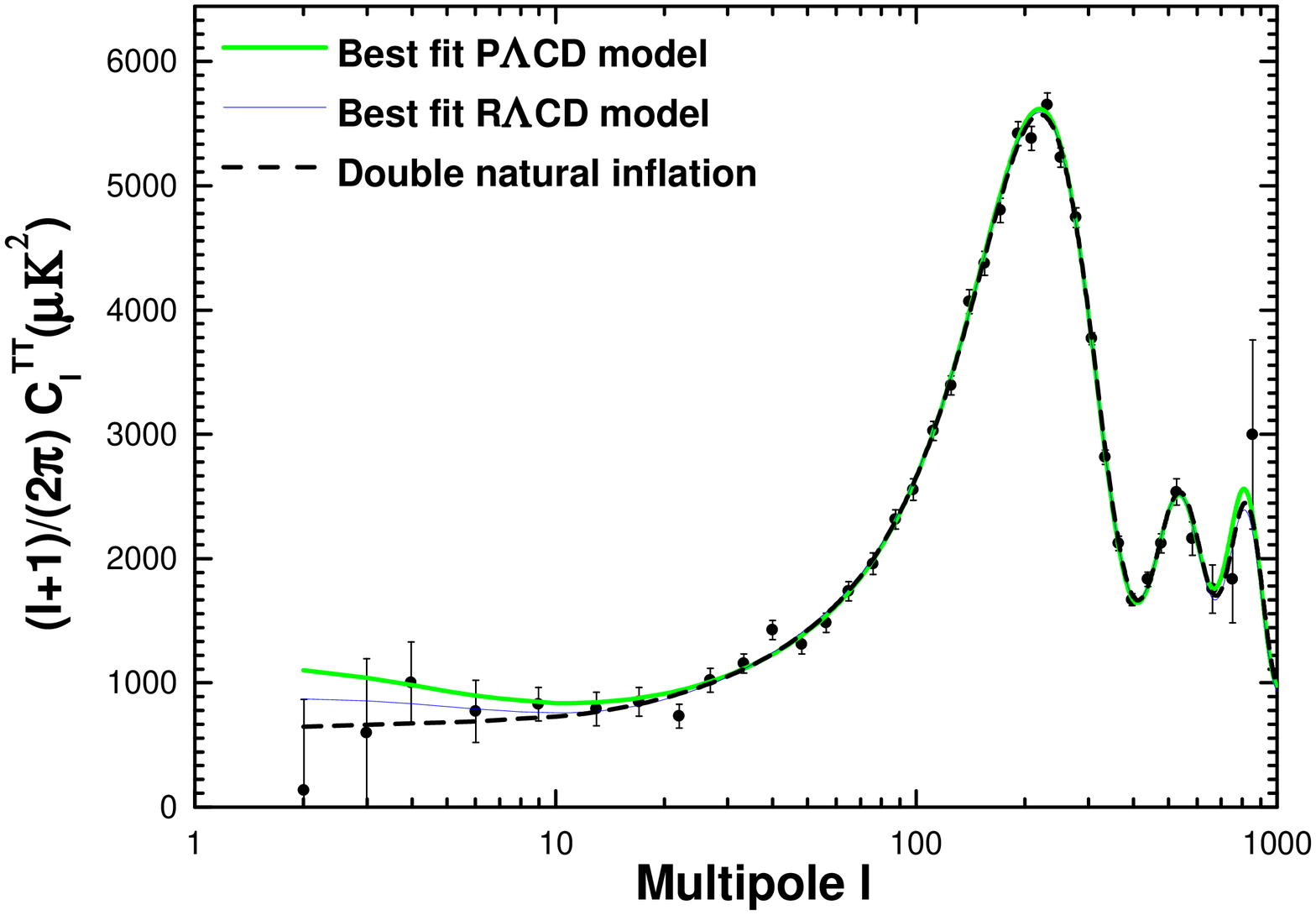}
\caption{Suppression of CMB quadrupole in the double natural
inflation model. The CMB data with error bars are taken from WMAP
collaborations\cite{WMAP1}. \label{fig:fig3}}
\end{center}
\end{figure}

Before concluding, we comment on the possible discrepancy between
predictions and COBE\cite{cobe}, WMAP observations on the largest
scales\cite{WMAP1,otzh}. As argued in Ref.~\onlinecite{Bridle},
the need of large running is due to the suppressed multipoles at
$l=2,3,4$. However as shown in Ref.~\onlinecite{WMAP1} the
possibility of finding a lower value of quadrupole in the presence
of a constant running of the spectral index is no more than 0.9
percent for a spacial flat $\Lambda$CDM  cosmology. In our model,
the running index $dn_s/d\ln k$ is not a constant, but we have
checked that it does not improve the fit significantly. With many
Abelian gauge fields in the higher dimensions, we will show now
this discrepancy can be alleviated. For simplicity of our
discussion, consider two Wilson lines $\theta_1$ and $\theta_2$,
the Lagrangian for such a system is given by
\bea\label{eqpotential2} {\cal L}&=&\frac{1}{2}(\partial\phi_1)^2
+\frac{1}{2}(\partial\phi_2)^2 -V_0\biggl[1-
\cos\left(\frac{q_1\phi_1}{f_{\rm
eff}}\right)\biggr]\nonumber\\
&-& V_0\biggl[1- \cos\left(\frac{q_2\phi_2}{f_{\rm
eff}}\right)\biggr] \,. \eea Assuming adiabatic process with
certain choice of parameters, {\it e.g.}, $q_1=5 q_2$, inflation
will be firstly driven by the heavy field $\phi_1$ and then by
$\phi_2$, with no interruption in between. The "double natural
inflation" can generate a primordial spectrum with a feature at
the largest scales favored by current data. In Fig.~\ref{fig:fig3}
we give an example with $q_1=5 $ and $q_2 =1$. The details of our
calculations is similar to Ref.~\onlinecite{smallL}: we
numerically calculate the primordial scalar and tensor spectra and
fit the resulting CMB TT and TE spectra using WMAP likelihood
code\cite{Verde} with CMBfast program \cite{cmbfast}. We compare
our result with WMAP team's best fit power law $\Lambda$CDM
(P$\Lambda$CDM) and running spectral index $\Lambda$CDM
(R$\Lambda$CDM) model. The CMB quadrupole as shown in
Fig.~\ref{fig:fig3} is better suppressed in this model.

In summary, the WMAP result of a varying spectral index, if
stands, could be used as a discriminator for inflationary models;
most extant models allow for only a tiny amount of scale
variations in the spectral index and could face a severe
challenge. We have studied in this paper the possibility of
building inflation models with large running spectral indices, and
specifically  proposed a higher-dimensional model which can
generate  $dn/d\ln k \sim -{\cal O}{(10^{-2})}$, favored by the
WMAP analyses.

{\bf{Acknowledgments:}}
This work was supported in part by National Natural
Science Foundation of China and by Ministry of
Science and Technology of China under Grant No. NKBRSF G19990754,
and in part by an NSF grant NSF-0070928.


\begin{thebibliography}{nn}


\bibitem{WMAP} C. L. Bennett {\it et al}, {\tt astro-ph/0302207}, to appear in
ApJ; also see the WMAP homepage, {\tt http://map.gsfc.nasa.gov}.

\bibitem{WMAP1}\label{ref:map1}
D. N. Spergel {\it et al.}, {\tt astro-ph/0302209}, to appear in
ApJ.

\bibitem{WMAP2}\label{ref:map2}
H. V. Peiris {\it et al.}, {\tt astro-ph/0302225}, to appear in
ApJ.

\bibitem{melchiorri}
A. Melchiorri and C.~J.~\"{O}dman, Phys. Rev. D {\bf 67}, 081302
(2003).

\bibitem{Bridle}
S. L. Bridle, A. M. Lewis, J. Weller, and G. Efstathiou, {\tt
astro-ph/0302306}, to appear in MNRAS.

\bibitem{Barger}
V. Barger, H.-S. Lee, and D. Marfatia, Phys. Lett. B {\bf 565}, 33
(2003).

\bibitem{SS}
B. Moore, Nature {\bf 370}, 629 (1994); R.~A.~Flores and
J.~A.~Primack, ApJ {\bf 427}, L1 (1994); W.~J.~G.~De Blok and
S.~S.~McCaugh, MNRAS {\bf 290}, 533 (1997); D. N. Spergel and P.
J. Steinhardt, Phys. Rev. Lett. {\bf 84}, 3760 (2000);
M.Kamionkowski and A. R. Liddle, Phys. Rev. Lett. {\bf 84}, 4525
(2000); A. R. Zentner and J. S. Bullock, Phys. Rev. D {\bf 66},
43003 (2002); W. Lin, D. Huang and X. Zhang, Phys. Rev. Lett. {\bf
86}, 954 (2001).

\bibitem{Liddle2}
See, {\it e.g.}, A. R. Liddle and D. H. Lyth, Cosmological
inflation and large-scale structure, Cambridge University Press (2000).

\bibitem{dodelson}
S. Dodelson and E. D. Stewart, Phys. Rev. D {\bf 65}, 101301 (2002).

\bibitem{stewart}
E. D. Stewart, Phys. Lett. B {\bf 391}, 34 (1997);
Phys. Rev. D {\bf 56}, 2019 (1997).

\bibitem{wang}
X. Wang {\it et al.}, {\tt astro-ph/0209242}.

\bibitem{copeland}
E. J. Copeland, A. R. Liddle, D. H. Lyth, E. D.
Stewart, and D. Wands, Phys. Rev. D {\bf 49}, 6410 (1994).

\bibitem{ACCR}
N.~Arkani-Hamed, H.-C.~Cheng, P.~Creminelli, and L.~Randall, Phys.
Rev. Lett. {\bf 90}, 221302 (2003); {\tt hep-th/0302034}, to
appear in JCAP.

\bibitem{KW}
D. E. Kaplan and N. Weiner, {\tt hep-ph/0302014}; M.~Fairbairn,
L.~L.~Honorez, and M.~H.~G.~Tytgat, Phys. Rev. D {\bf 67}, 101302
(2003).

\bibitem{Freese}
K.~Freese, J.~A.~Frieman, and A.~V.~Olinto, Phys.\ Rev.\ Lett.\
{\bf 65}, 3233 (1990); F.~Adams, J.~R.~Bond, K.~Freese,
J.~A.~Frieman, and A.~V.~Olinto, Phys.\ Rev.\ D {\bf 47}, 426
(1993).

\bibitem{hosotani}
Y. Hosotani, Phys. Lett. B {\bf 126}, 309 (1983);
{\bf 129}, 193 (1983); Ann. Phys. {\bf 190}, 233 (1989).

\bibitem{Delgado}
A. Delgado, A. Pomarol, and M. Quiros, Phys. Rev. D {\bf 60},
095008 (1999).

\bibitem{largez}
S. Dimopoulos and S. Thomas, {\tt hep-th/0307004}.

\bibitem{prepare}
B. Feng, M. Li and X. Zhang, in preparation.

\bibitem{riotto}
L.~Pilo, D.~A.~J.~Rayner, and A.~Riotto, {\tt hep-ph/0302087}.

\bibitem{feng} B. Feng, X. Gong, and X. Wang, {\tt
astro-ph/0301111}.

\bibitem{liddle}
A.~R.~Liddle, A.~Mazumdar, and F.~E.~Schunck, Phys.~Rev.~D {\bf
58}, 061301 (1998).

\bibitem{cobe}
C.~L.~Bennett {\it et al.}, Astrophys.\ J.\  {\bf 464}, L1 (1996).

\bibitem{otzh}
A. Oliveira-Costa, M. Tegmark, M. Zaldarriaga and A. Hamilton,
{\tt astro-ph/0307282}.

\bibitem{smallL}
Bo~Feng and Xinmin~Zhang, {\tt astro-ph/0305020}.

\bibitem{Verde}
L.~Verde {\it et al.}, {\tt astro-ph/0302218}, to appear in ApJ.

\bibitem{cmbfast}
U.~Seljak and M.~Zaldarriaga, Astrophys.\ J.\  {\bf 469}, 437
(1996).


\end{thebibliography}
\end{document}